\documentstyle[12pt]{article}
\input epsf
\begin{document}

\title{Unitary Transformation in Probabilistic Teleportation}
\author{Xiu-Lao Tian\footnote{Corresponding author: E-mail, txl@xupt.edu.cn},
        ~~Wei Zhang,~~Xiao-Qiang Xi\\
        {\small School of Science, Xi'an University of Posts and
        Telecommunications, Xi'an 710121, China}}

\date{\today}
\maketitle

\begin{abstract}
We proposed a general transformation in probabilistic teleportation, which
is based on different entanglement matching coefficients $K$ corresponding
to different unitary evolution which provides one with more flexible
evolution method experimentally. Through analysis based on the Bell
basis and generalized Bell basis measurement for two probabilistic
teleportation, we suggested a general probability of successful
teleportation, which is not only determined by the entanglement
degree of transmission channels and measurement methods, but also
related to the unitary transformation in the teleportation process.
\end{abstract}



\maketitle

\section{Introduction}

Quantum entanglement is one of the most fascinating characteristic of
quantum physics, a fantastic application of entanglement
\cite{R Horodecki2009,PRA74 052105,PRA79 024303} is quantum teleportation,
which plays a key role in the field of quantum communication. Since
the seminal work of Bennett et al. \cite{Bennett1993}, teleportation
has been the research interest of researchers and a number of work
both in theory and experiments has been devoted to it
\cite{Bou1997,Nielsen1998,JOP2004,PRL Y. Yeo2006,Nature Lett2008,
PLA375 922,PLA375 2140}.

Up to now the teleportation has been studied in different branches,
such as directly and network controlled teleportation \cite{PRA
Karlsson1998,PRA F G Deng 2005,PRA X Z Man2007,OPC S G F2009};
discrete-variables and continuous-variables teleportation \cite{PRA
CV Te 1998,PRA P M2006,IJQI P M2008}; prefect and probabilistic
teleportation \cite{PRA  W L Li2000,PLA B S Shi2000} and so on. In
fact, one of the key problem of teleportation is how to construct an
usefulness quantum channel, different channels will yield different
results, some channels can be used to realize perfect teleportation,
while some others can only enable probabilistic teleportation. Because
of the inevitable interaction with its surroundings, correlations in
quantum states are difficult to maintain \cite{PLA374 3520,AOP},
therefore the probabilistic teleportation \cite{CPL H Lu2001,PRA J
Fang2003,CPL T Gao2003,CPL F L Yan2006,CPL W X Jiang2007} has been
widely discussed in recent years.

A necessary and sufficient condition for realizing perfect
teleportation and successful teleportation has been given in
\cite{T1,T2,T3,T4}. Based on the Bell basis measurement, we
found that if the channel parameter matrix (CPM) is unitary,
then one can always realize a perfect teleportation (i.e.,
the successful probability $p=1$), if the CPM is invertible but not
unitary, however, one can only realize a probabilistic teleportation
(i.e., the successful probability $p<1$).

Motivated by the idea of Ref. \cite{PRA W L Li2000}, in which the
authors introduced an auxiliary qubit to realize probabilistic
teleportation, we propose some unitary transformation methods
in probabilistic teleportation in this work. These methods are based
on different entanglement matching coefficients $K$ which corresponding
to different unitary transformations. Through detailed analysis with
the Bell basis and generalized Bell basis measurement in two
probabilistic teleportation processes, we suggest a general
probability of successful teleportation, which is not only
determined by the entanglement degree of transmission channel and
measurement methods, but also related to the unitary transformation
in the teleportation process. For example, if we teleport the
unknown one-qubit state $|\varphi\rangle_{1}$ via the channel state
$|\varphi\rangle_{2,3}=a|00\rangle +b|11\rangle$, then the whole
probability of successful teleportation is $P= 2(Kab)^2$, where
$0<K\leq \min(\frac{1}{|a|},\frac{1}{|b|})$. Our conclusion covers
and complements the results of Ref. \cite{PRA  W L Li2000}. As different
$K$ will give different kinds of evolution methods, one
can have more flexible and selectable evolution method
experimentally.

\section{Entanglement matching and probabilistic teleportation}
Suppose Alice wants to send an unknown one-qubit state
$|\varphi\rangle_{1}$ to Bob
\begin{eqnarray}
 |\varphi\rangle_{1}=R^{i}|i\rangle
 =R^{0}|0\rangle+R^{1}|1\rangle=\alpha|0\rangle+\beta|1\rangle,
\end{eqnarray}
where $R^{i}R_{i}^*=|\alpha|^2+|\beta|^2=1$, with $i$ being
taken to be 0 or 1 and a repeated index denotes summation.

The general two-qubit state $|\varphi\rangle_{2,3} $ used as the quantum
channel can be expressed as follows
\begin{eqnarray}
 |\varphi\rangle_{2,3}=\frac{1}{\sqrt{2}}X^{jk}|jk\rangle=
 \frac{1}{\sqrt{2}}(X^{00}|00\rangle
 +X^{01}|01\rangle +X^{10}|10\rangle+X^{11}|11\rangle),
\end{eqnarray}
where $X^{jk}X_{jk}^*=2$. If Alice adopts the standard Bell basis
measurement (BM) $\phi_{ij}^{\lambda}$ $(\lambda=1,2,3,4)$
on her particles, i.e., $\phi_{ij}^{1,2}=(|00\rangle
\pm|11\rangle)/ \sqrt{2}$ and
$\phi_{ij}^{3,4}=(|01\rangle \pm |10\rangle)/ \sqrt{2}$.
Then with the denotation of Bell basis \cite{T1,T2,T3}, the
total state of the system can be rewritten as
\begin{eqnarray}
 |\Psi\rangle_{tot}=\frac{1}{\sqrt{2}}R^{i}X^{jk}
 {L}^\lambda_{ij}|\lambda k\rangle
 =\frac{1}{{2}}R^{i}X^{jk}{T}^\alpha_{ij}|\alpha k\rangle
 =\frac{1}{2}R^{i}\sigma^{(\lambda)k}_{i}|\alpha k\rangle,
\end{eqnarray}
where $\sigma_i^{(\lambda)k}=X^{jk}T_{ij}^{\lambda}=X^{jk}{\sqrt{2}}
{L}^\alpha_{ij}$ is the element of
\begin{eqnarray}
 \sigma^{\alpha}=XT^{\alpha}=
 \left(\begin{array}{cc}
 \sigma_{0}^{\lambda0}&\sigma_{1}^{\lambda0}\\
 \sigma_{0}^{\lambda1}&\sigma_{1}^{\lambda1}
 \end{array}\right)=
 \left(\begin{array}{cc}
 X^{00}&X^{10}\\
 X^{01}&X^{11}
\end{array}\right)
\left(\begin{array}{cc}
 T_{00}^{\lambda}&T_{10}^{\lambda}\\
 T_{01}^{\lambda}&T_{11}^{\lambda}
\end{array}\right).
\label{sigmat}
\end{eqnarray}

After Alice's measurement, the total state will collapse to
\begin{eqnarray}
 |\Psi^{\alpha}\rangle_B=\frac{1}{2}R^{i}{\sigma}_i^{(\alpha)k}|k\rangle.
\end{eqnarray}
Obviously, based on the BM method, all the $T^\alpha$
are unitary. So if the CPM $X$ is unitary, one can always realize
the perfect teleportation (i.e., the whole probability $p=1$), if
$X$ is invertible but not unitary, one can only realize a
probabilistic teleportation (i.e., the whole probability $p<1$).

In Ref. \cite{PRA  W L Li2000}. Li et al. presented a protocol
of probabilistic teleportation by introducing an auxiliary qubit state
$|0\rangle_{A}$ and performing an unitary transformation on Bob's
state. They employ a partially entangled state as the quantum channel,
that is
\begin{eqnarray}
 |\varphi\rangle_{2,3}=X^{jk}|jk\rangle =a|00\rangle
 +b|11\rangle~~(a\neq b),
\end{eqnarray}
with $a, b$ being real numbers and $a^2+b^2=1$. The CPM
$X=\sqrt{2}{\rm diag}(a,b)$ is obviously invertible but not unitary, so Bob
cannot directly retrieve the state by acting $(\sigma^{\alpha})^{-1}$
on the collapsed state $|\Psi^{\alpha}\rangle_B$.

With the standard Bell basis $\phi^\lambda_{ij}~(\lambda=1,2,3,4)$,
the total state of the system can be rewritten as
\begin{eqnarray}
|\Psi\rangle_{tot}&=&\frac{1}{\sqrt{2}}R^{i}X^{jk}T^\lambda_{ij}|\lambda
                      k\rangle =\frac{1}{\sqrt{2}}R^{i}\sigma^{(\lambda)k}_{i}|\lambda
                      k\rangle\nonumber\\
                  &=&\frac{1}{\sqrt{2}}[\phi^1_{1,2}(a\alpha|0\rangle+b\beta|1\rangle)
                      +\phi^2_{1,2}(a\alpha|0\rangle-b\beta|1\rangle)\nonumber\\
                  &&+\phi^3_{1,2}(a\beta|0\rangle+b\alpha1\rangle)\
                    +\phi^4_{1,2}(a\beta|0\rangle-b\alpha|1\rangle)].
\end{eqnarray}

After Alice's BM $\phi^1_{12}$, Bob will get
the unnormalized state as follows
\begin{eqnarray}
|\Psi^1\rangle_{b}=\frac{1}{\sqrt{2}}R^{i}\sigma^{(1)k}_{i}|
k\rangle
=\frac{1}{\sqrt{2}}(a\alpha|0\rangle+b\beta|1\rangle).
\end{eqnarray}

In order to obtain the original state, one can introduce an
auxiliary qubit state $|0\rangle_{A}$ \cite{PRA  W L Li2000} in
Bob's state, now Bob's state can be express as
\begin{eqnarray}
 |\Psi^1\rangle_{A,3}=\frac{1}{\sqrt{2}}R^{i}\sigma^{(1)k}_{i}|
 k\rangle
 =\frac{1}{\sqrt{2}}|0\rangle_{A}[(a\alpha|0\rangle+
 b\beta|1\rangle)_{3},
\end{eqnarray}

With the standard two-qubit basis $(|00\rangle, |01\rangle, |10\rangle,
|11\rangle)_{A,3}$,  we propose an unitary transformation $U$ with
parameter $K$ for the particles (A,3)
\begin{eqnarray}
 U=\left(\begin{array}{cccc}
 Kb&0&\sqrt{1-(Kb)^2}&0\\
 0&Ka&0&\sqrt{1-(Ka)^2}\\
 \sqrt{1-(Kb)^2}&0&-Kb&0 \\
 0&\sqrt{1-(Ka)^2}&0&-Ka\\
 \end{array}\right).
\end{eqnarray}
where we called $K$ the entanglement matching coefficient of Bob's
evolution. To ensure the transformation $U$ to be
unitary, we demand $0<K\leq \min(\frac{1}{|a|},\frac{1}{|b|}) $.
There are different unitary transformation methods with different $K$.

After Bob's evolution, the state $|\Psi^1\rangle_{A,3}$ turns out to be
\begin{eqnarray}
|\Psi^1\rangle_{A,3}=\frac{1}{\sqrt{2}}[Kab
|0\rangle_{A}(\alpha|0\rangle+\beta|1\rangle)_{3}
+a{\sqrt{1-(Kb)^2}}\alpha|1\rangle_{A}|0\rangle_{3}
+b{\sqrt{1-(Ka)^2}}\beta|1\rangle_{A}|1\rangle_{3}.
\end{eqnarray}

Certainly, $|\Psi^1\rangle_{A,3} $ is not normalized. Now Bob
performs measurement on the auxiliary qubit $A$, if
the measurement outcome is $ |1\rangle_{A}$, the teleportaton fails,
if the measurement outcome is $ |0\rangle_{A}$, the teleportation is
successfully accessed and Bob's state becomes
\begin{eqnarray}
|\Psi^1\rangle_{3}=\frac{1}{\sqrt{2}}Kab(\alpha|0\rangle+\beta|1\rangle)_{3}
\end{eqnarray}

Now we discuss the probability of successful teleportation, which contains
both Alice's $P_A$ and Bob's $P_B$ probability.
From Eq. (7) one can obtain the Bell state $\phi^1_{1,2}$ occurring
probability as
\begin{eqnarray}
P^1_A=\frac{1}{{2}}\langle0|a\alpha\langle1|b\beta
(|a\alpha|0\rangle+b\beta|1\rangle)_{3}=\frac{1}{{2}}[(a\alpha)^2+(b\beta)^2],
\end{eqnarray}
Similarly, the Bell state $\phi^2_{1,2}$, $\phi^3_{1,2}$ and
$\phi^4_{1,2}$ occurring probability are
\begin{eqnarray}
P^2_A=P^1_A=\frac{1}{{2}}[(a\alpha)^2+(b\beta)^2],
P^3_A=P^4_A=\frac{1}{{2}}[(a\beta)^2+(b\alpha)^2],
\end{eqnarray}
If $a=b=\frac{1}{\sqrt{2}},$ then
$P^1_A=P^2_A=P^3_A=P^4_A=\frac{1}{{4}}$, which is just the prefect
teleportation.

Now we compute the probability of Bob for obtaining
the original state from the state $|\Psi^1\rangle_{A,3}$. The normalized
state corresponding to the state in Eq. (11) is
\begin{eqnarray}
|\Psi^1\rangle_{(A,3)norm}=\frac{1}{\sqrt{a^2\alpha^2+b^2\beta^2}}|\Psi^1\rangle_{A,3}.
\end{eqnarray}

After Bob's successful measurement on $ |0\rangle_{A}$, the
probability $P^1_B$ of obtaining the original state from the
state $|\Psi^1\rangle_{A,3}$ is
\begin{eqnarray}
P^1_B=\frac{(Kab)^2}{{(a\alpha)^2+(b\beta)^2}}.
\end{eqnarray}

We consider Alice's measurement and Bob's different operations in the
teleportation process. For Alice's each measurement and Bob's operation,
the probability of obtaining the initial state is
\begin{eqnarray}
P^1_{AB}=P^1_AP^1_B=\frac{1}{2}(Kab)^2
\end{eqnarray}
where $P^1_{AB}$ is just the square of coefficient of the state
$|\Psi^1\rangle_{3}$ in Eq. (12).

Summing all the contributions of $
P^1_{AB}=P^2_{AB}=P^3_{AB}=P^4_{AB}=\frac{1}{2}(Kab)^2$, we obtain the whole
probability of successful teleportation as
\begin{eqnarray}
P=\sum_{i=1}^4=2(Kab)^2
\end{eqnarray}
where $0<K\leq \min(\frac{1}{|a|},\frac{1}{|b|}) $.

If $a>b $, we take $K=K_{max}=\frac{1}{a}$,
then the optimal probability is
$ P=2b^2$. When $a=b=\frac{1}{\sqrt{2}},$ then $P_{max}=1$, which is
just the case for prefect teleportation, and for this special case $U$
is the same as that in Ref. \cite{PRA  W L Li2000}, i.e.
\begin{eqnarray}
U=\left(\begin{array}{cccc}
   b/a&0&\sqrt{1-(b/a)^2}&0\\
   0&1&0&0\\
   \sqrt{1-(b/a)^2}&0&-b/a&0 \\
   0&0&0&-1\\
\end{array}\right).
\end{eqnarray}

For different matching coefficients $K$, one can adopt different kinds of
unitary transformation. FOr example, when $K=1$ the
whole probability of successful teleportation is $P=2(ab)^2$ and our
unitary transformation matrix turns out to be
\begin{eqnarray}
U=\left(\begin{array}{cccc}
b&0&a&0\\
0&a&0&b\\
a&0&-b&0 \\
0&b&0&-a\\
\end{array}\right).
\end{eqnarray}

Next we discuss some difference between the probability $P=2b^2$ and
$P=2(Kab)^2$ for two kinds of unitary transformation. For arbitrary
$a^2+b^2=1$ and $a\neq b$, there are always $P=2b^2>2(ab)^2$. So
Eq. (20) is an unitary transformation for obtaining the optimal
probability. However, the condition $a^2+b^2=1$ and $a>b$ yields
$b^2<1/2$, and therefore one can only obtain probability $P<1$.
When $K=1$ we have $P=2(ab)^2$, and the normalization condition
$a^2+b^2=1$ gives rise to $P_{max}=1/2$ with $a^2=b^2=1/2$, thus
one can only attain the probability $P=2(ab)^2<1/2$ for $a\neq b$.

Because $0<K\leq \min(\frac{1}{|a|},\frac{1}{|b|})$, when $1\leq K\leq 2$
(here $a>b$ and $a\sim b$), then $ P=4(ab)^2\sim 2b^2$. For this case
there are a little difference between the probability $P=2b^2$ and
$P=4(ab)^2$ for the two kinds of unitary transformation (see Fig. 1).
\begin{center}
\epsfxsize 90mm \epsfysize 40mm \epsffile{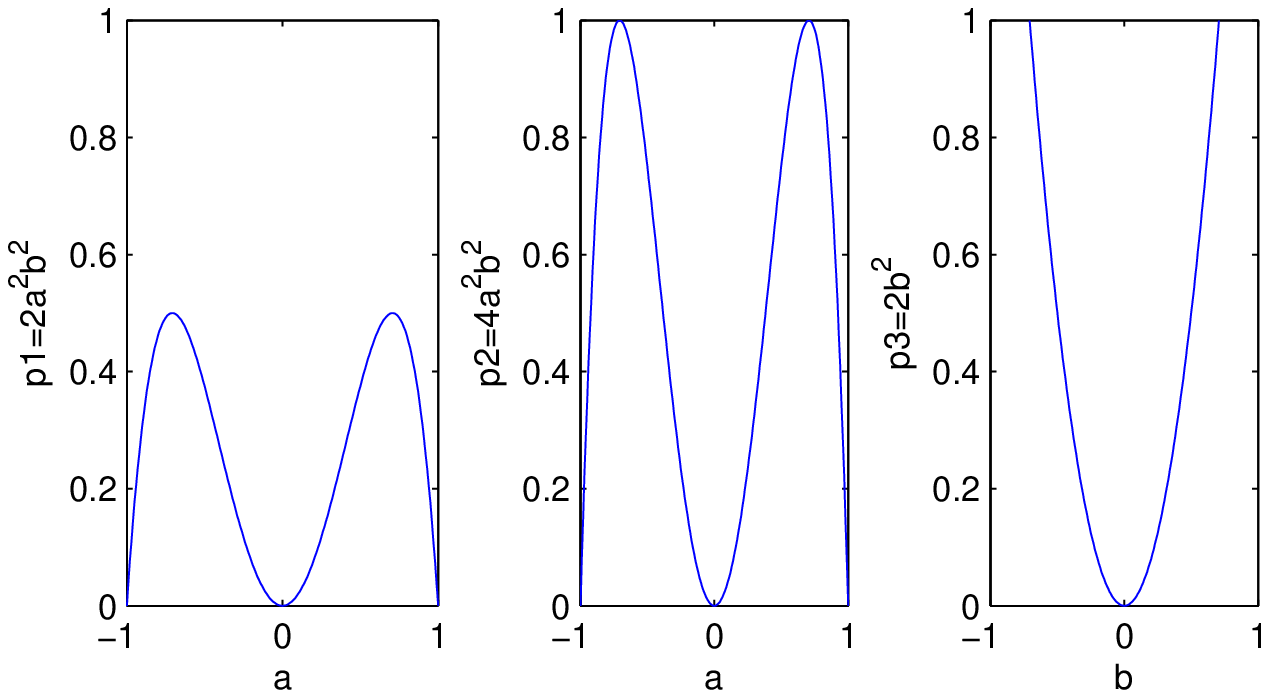}

{\footnotesize Fig. 1. Comparison of the probability with different channel
parameters.}
\end{center}

Different matching coefficients $K$ correspond to different
unitary transformations. Although $P=2b^2>2(Kab)^2$, it provides
one with more flexible selection for probabilistic teleportation.

\section{Probabilistic teleportation with generalized BM}

Considering now Alice makes a generalized Bell basis measurement (GBM), these
are
\begin{eqnarray}
\phi^1_{1,2}=a'|00\rangle+b'|11\rangle,~~
\phi^2_{1,2}=b'|00\rangle-a'|11\rangle,~~
\phi^3_{1,2}=a'|01\rangle+b'|10\rangle,~~
\phi^4_{1,2}=b'|01\rangle-a'|11\rangle.
\end{eqnarray}
where $a'^{2}+b'^2=1$ and $a'\neq b'.$ The transformation matrix ${T}$
between the generalized Bell basis and
computation basis $\{|00\rangle, |01\rangle, |10\rangle, 11\rangle\}$ is
\begin{eqnarray}
{T}= \left(\begin{array}{cccc}
a'&0&0&b' \\
b'&0&0&-a' \\
0&a'&b'&0 \\
0&b'&-a'&0
 \end{array}\right).
 \end{eqnarray}

Let us reconsider the aforementioned one-qubit teleportation, under the
generalized Bell basis, the total state of the system is
\begin{eqnarray}
|\Psi\rangle_{tot}=\frac{1}{\sqrt{2}}R^{i}X^{jk}T^\lambda_{ij}|\lambda
k\rangle =\frac{1}{2}R^{i}\sigma^{(\lambda)k}_{i}|\lambda k\rangle.
\end{eqnarray}

After Alice's GBM $ \phi^\lambda_{12}$, Bob will get the corresponding
 unnormalized state as follows
\begin{eqnarray}
&&|\Psi^1\rangle_B=(aa'\alpha|0\rangle+bb'\beta|1\rangle),~~
  |\Psi^2\rangle_B=(ab'\alpha|0\rangle-ba'\beta|1\rangle),\nonumber\\
&&|\Psi^3\rangle_B=(aa'\beta|0\rangle+bb'\alpha|1\rangle),~~
  |\Psi^4\rangle_B=(ab'\beta|0\rangle-ba'\alpha|1\rangle).
\end{eqnarray}

When Bob introduced an auxiliary qubit state $|0\rangle_{A}$ for
$|\Psi^\alpha\rangle_B$, and make an unitary transformation
$U_1$ to the state $|\Psi^{1}\rangle_{A,3}$, with
\begin{eqnarray}
U_1=\left(\begin{array}{cccc}
Kbb'&0&\sqrt{1-(Kbb')^2}&0\\
0&Kaa'&0&\sqrt{1-(Kaa')^2}\\
\sqrt{1-(Kbb')^2}&0&-Kbb'&0 \\
0&\sqrt{1-(Kaa')^2}&0&-Kaa'\\
\end{array}\right),
\end{eqnarray}
where $0<K\leq \min(\frac{1}{|aa'|},\frac{1}{|bb'|})$, the state
$|\Psi^1\rangle_{A,3}$ becomes
\begin{eqnarray}
|\Psi^1\rangle_{A,3}&=&\frac{1}{\sqrt{2}}|0\rangle_{A}[Kaba'b'(\alpha|0\rangle+\beta|1\rangle)_{3}\nonumber\\
&+&a{\sqrt{1-(Kbb')^2}}\alpha|1\rangle_{A}|0\rangle_{3}\
+b{\sqrt{1-(Kaa')^2}}\beta|1\rangle_{A}|1\rangle_{3}.
\end{eqnarray}

If Bob's measurement outcome is $|0\rangle_{A}$, the teleportation
is successfully implemented.

Considering Alice's Bell basis measurement $\phi^1$ and Bob's
evolution on $|\Psi^1\rangle_{A,3}$,
the probability of successful teleportation is
\begin{eqnarray}
P^{1}_{AB}=(Kaba'b')^2.
\end{eqnarray}
Similarly, After Bob's $U_2$, $U_3$ and $U_4$ transformation to the
corresponding states and measurement on the auxiliary qubit, the
probability of successful teleportation respectively are
\begin{eqnarray}
P^2_{AB}=P^3_{AB}=P^4_{AB}=(Kaba'b')^2.
\end{eqnarray}
Thus the whole probability of successful teleportation is
\begin{eqnarray}
 P=4(Kaba'b')^2.
\end{eqnarray}

Next we discuss the optimal probability of successful teleportation
by entanglement matching for the following two cases:

(1) $|a|\geq|a'|\geq|b'|\geq|b|$. For this case, $|aa'|\geq|bb'|$,
$|ab'|\geq|ba'|$, so we take $K=\frac{1}{|aa'|}$ in $U_1$ and $U_3$,
$K=\frac{1}{|ab'|}$ in $U_2$ and $U_4$, for which one can obtain
$P^1_{AB}=P^3_{AB}=|bb'|^2$, $P^2_{AB}=P^4_{AB}=|ba'|^2$. The optimal
whole probability is
\begin{eqnarray}
 P=\sum p_i=|bb'|^2+|ba'|^2+|bc'|^2+|bd'|^2=2|b|^2.
\end{eqnarray}

(2) $|a'|\geq|a|\geq|b|\geq|b'|$. In this case, $|aa'|\geq|bb'|$,
$|ba'|\geq|ab'|$, so we take $K_1=\frac{1}{|aa'|}$, $K_2=\frac{1}{|ba'|}$,
for which we have $P^1_{AB}=P^3_{AB}=|bb'|^2$,
$P^2_{AB}=P^4_{AB}=|ab'|^2$. The optimal whole probability is
\begin{eqnarray}
 P=\sum p_i=2|b'|^2.
\end{eqnarray}

From the above analysis, one can see that the optimal probability of
successful teleportation is determined by the smaller value of $|b|$
and $|b'|$, i.e., the optimal probability is determined by the
entanglement degree of Alice's measurement or the quantum channel.
However, for $K<K_{max}=\min(\frac{1}{|aa'|},\frac{1}{|bb'|})$, the
whole probability of successful teleportation is $P=4(Kaba'b')^2$.
Therefor, the general probability of successful teleportation is not only
determined by the factors of the channel and measurement, but also related
to the unitary transformation during teleportation process.

\section{Conclusion}
The unavoidable influence of environment always induces degradation of
quantum correlations, therefor the study of probabilistic teleportation
is significant for quantum information processing. In this paper, we
generalized the protocol of probabilistic teleportation by introducing
an auxiliary qubit and the unitary transformation methods.
Moreover, through the analysis based on the Bell basis and generalized
Bell basis measurement in two probabilistic teleportation, we suggested
a general probability of successful teleportation, which is not only
determined by both the entanglement degree of transmission channels and
the measurement methods, but also related to unitary transformation in
teleportation process, i.e., $P=2(Kab)^2$, Although $ P=2(Kab)^2<2(b)^2$
(the optimal $U$ transformation). However in experiment, it is more important
to realize successful teleportation. As different entanglement matching
coefficients $K$ will give different $U$ evolution methods, so one can have
more flexible selectable evolution method experimentally.
\\

\textbf{Acknowledgments}\\
\indent This work was supported in part by the National Natural Science Foundation of China under
Grant No. 10902083, the Natural Science Foundation of Shaanxi Province under Grant Nos.
2009JQ8006, 2009JM6001 and 2010JM1011.


\begin{thebibliography}{36}

\bibitem{R Horodecki2009} R. Horodecki, P. Horodecki, M. Horodecki, K. Horodecki, {Rev. Mod. Phys.} 81 (2009) 865.
\bibitem{PRA74 052105}A. Abliz, H. J. Gao, X. C. Xie, Y. S. Wu, W. M. Liu, {Phys. Rev. A} 74 (2006) 052105.
\bibitem{PRA79 024303}Z. G. Li, S. M. Fei, Z. D. Wang, W. M. Liu, {Phys. Rev. A} 79 (2009) 024303.
\bibitem{Bennett1993}C. H. Bennett, G. Brassard, C. Cr\'{e}peau {\em et al.}, {Phys. Rev. Lett.} 70 (1993) 1895.
\bibitem{Bou1997} D. Bouwmeester, J. W. Pan, K. Mattle {\em et al.}, {Nature} 390 (1997) 575.

\bibitem{Nielsen1998} M. A. Nielsen, E. Knill, R. Laflamme, {Nature} 396 (1998) 52.
\bibitem{JOP2004}H. Y. Dai, P. X. Chen, C. Z. Li, {J. Opt. B: Quantum Semiclass. Opt.} {6} (2004) 106.
\bibitem{PRL Y. Yeo2006}Y. Yeo, W. K. Chua, {Phys. Rev. Lett.} 96 (2006) 060502.
\bibitem{Nature Lett2008} Z. S. Yuan, {\em et al.}, {Nature Phys} {454} (2008) 1098.
\bibitem{PLA375 922}M. L. Hu, Phys. Lett. A 375 (2011) 922.

\bibitem{PLA375 2140}M. L. Hu, Phys. Lett. A 375 (2011) 2140.
\bibitem{PRA Karlsson1998}A. Karlsson, M. Bourennane, {Phys. Rev. A} {58} (1998) 4394 .
\bibitem{PRA F G Deng 2005} F. G. Deng, {\em et al.},  {Phys. Rev. A} {72} (2005) 022338.
\bibitem{PRA X Z Man2007} Z. X. Man, Y. J. Xia, N. B. An, {Phys. Rev. A} {75} (2007) 052306.
\bibitem{OPC S G F2009} G. F. Shi, X. Q. Xi, X. L. Tian, R. H. Yue, {Opt. Commun.} {282} (2009) 2460.

\bibitem{PRA CV Te 1998}S. L. Braunstein, H. J. Kimble, {Phys. Rev. Lett.} 80 (1998) 869.
\bibitem{PRA P M2006}P. Marian, T. A. Marian, {Phys. Rev. A} 74 (2006) 042306.
\bibitem{IJQI P M2008}P. Marian, T. A. Marian, {Int. J. Quantum Inf.} {6} (2008) 721.
\bibitem{PRA  W L Li2000} W. L. Li, C. F. Li, G. C. Guo, {Phys. Rev. A} {61} (2000) 34301.
\bibitem{PLA B S Shi2000}B. S. Shi, Y. K. Jiang, G. C. Guo, {Phys. Lett. A} {268} (2000) 161.

\bibitem{PLA374 3520}M. L. Hu, Phys. Lett. A 374 (2010) 3520.
\bibitem{AOP}M. L. Hu, H. Fan, Ann. Phys. 327 (2012) 851.
\bibitem{CPL H Lu2001}H. Lu, {Chin. Phys. Lett.} {18} (2001) 1004.
\bibitem{PRA J Fang2003} J. Fang, Y. Lin, S. Zhu, X. Chen, {Phys. Rev. A} {67} (2003) 014305.
\bibitem{CPL T Gao2003} T. Gao, Z. X. Wang, F. L. Yan, {Chin. Phys. Lett.} {20} (2003) 2094.

\bibitem{CPL F L Yan2006} F. L. Yan, H. W. Ding, {Chin. Phys. Lett.} {23}, 17 (2006).
\bibitem{CPL W X Jiang2007}W. X. Jiang, {\em et al.} {Chin. Phys. Lett} {24} (2007) 1144.
\bibitem{T1} X. L. Tian, X. Q. Xi, {Int. J. Quantum Inform.} {7} (2009) 927.
\bibitem{T2} X. L. Tian, {\em et al.}, {Opt. Commun.} {282} (2009) 4815.
\bibitem{T3} X. L. Tian, X. Q. Xi, {Mod. Phys. Lett. B} {23} (2009) 2261.

\bibitem{T4} X. L. Tian, M. L. Hu, X. Q. Xi, {Int. J. Theor. Phys.} {48} (2009) 2610.


\end{thebibliography}
\end{document}